\documentclass[prb,aps,superscriptaddress,floatfix,twocolumn]{revtex4}
\usepackage{graphicx,amssymb,amsmath}
\usepackage{epstopdf}
\begin{document}

\title{$1/f$ spectrum and memory function analysis of solvation dynamics in a room-temperature ionic liquid}

\author{Daun Jeong}
\affiliation{Department of Chemistry, Seoul National University, Seoul 151-747, Korea}

\author{M.\ Y.\ Choi}
\affiliation{Department of Physics and Astronomy, Seoul National University, Seoul 151-747, Korea}
\affiliation{Asia-Pacific Center for Theoretical Physics, Pohang University of Science and Technology,
Pohang 790-784, Korea}

\author{YounJoon Jung}
\altaffiliation{Author to whom correspondence should be addressed. E-mail: yjjung@snu.ac.kr}
\affiliation{Department of Chemistry, Seoul National University,
Seoul 151-747, Korea}

\author{Hyung J.\ Kim}
\affiliation{Department of Chemistry, Carnegie Mellon University, Pittsburgh, PA 15213, U.S.A.}

\date{\today}

\begin{abstract}

To understand the non-exponential relaxation associated with solvation dynamics in the ionic liquid 1-ethyl-3-methylimidazolium hexafluorophosphate, we study power spectra of the fluctuating Franck-Condon energy gap of a diatomic probe solute via molecular dynamics simulations. Results show $1$/{\it f} dependence in a wide frequency range over 2 to 3 decades, indicating distributed relaxation times. We analyze the memory function and solvation time in the framework of the generalized Langevin equation using a simple model description for the power spectrum.  It is found that  the crossover frequency toward the white noise plateau is directly related to the time scale for the memory function and thus the solvation time.  Specifically, the low crossover frequency observed in the ionic liquid leads to a slowly-decaying tail in its memory function and long solvation time. By contrast, acetonitrile characterized by a high crossover frequency and (near) absence of $1/f$ behavior in its power spectra shows fast relaxation of the memory function and single-exponential decay of solvation dynamics in the long-time regime.
\end{abstract}

\maketitle

\section{Introduction}

Solvation dynamics in room-temperature ionic liquids (RTILs) have received intensive theoretical\cite{shim:letter,margulis:rtil:dyn,kobrak:rtil:dyn1,shim2,bhargava:rtil,kobrak:rtil:dyn2,jeong:pol,shim:rtil:review,kobrak:rtil:dyn3}  and experimental attention recently.\cite{samanta:rtil,maroncelli:rtil,sarkar:rtil,petrich:rtil,vauthey:rtil:dyn,castner:il,review}
The collective influence of the solvent in the presence of an optically-active probe solute is usually monitored via various dynamic electronic spectroscopies and described in terms of time-dependent fluctuations and relaxation of the Franck-Condon (FC) transition energy of the solute in solution. Often observed in RTILs is biphasic relaxation comprised of ultrafast sub-picosecond dynamics and ensuing non-exponential decay. Short-time solvation dynamics, arising mainly from small-amplitude inertial translational motions of solvent ions, make a substantial contribution to overall solvent relaxation despite their high viscosity,\cite{shim2,kobrak:rtil:dyn2} and thus can play an important role in reaction dynamics\cite{kim:shim:rtil:et} in RTILs. The subsequent relaxation, attributed to diffusive dynamics, involves cooperative movement of ions and accompanying structural relaxation.

A spectral analysis of the FC energy gap, which is equivalent to analyzing its time correlation function, is useful for investigating dynamics over various time scales. Of particular interest are long-time fluctuations that often result in
non-exponential relaxation.  For instance, $1/f$ behavior for long-time fluctuations in water\cite{Ramaswamy:water:1/f} and liquid silica\cite{Chakravarty:1/f} has been reported. Defect fluctuations in a disordered two-dimensional liquid also exhibit $1/f$ spectra, suggesting that system dynamics are  heterogeneous.\cite{Reichhardt:defect:1/f}
Slow relaxation in these systems indicates that the structural memory persists for a long time in spite of the fast inertial motion of solvent molecules.

In the generalized Langevin equation (GLE) description of solvation dynamics,\cite{solventfreq,hynes:GLE} the  time-dependent friction plays the role of a memory function. The temporal behavior of the memory function for a variety of liquid systems has been analyzed via numerical transforms of appropriate time correlation functions\cite{shim2,Egelstaff:memory,solventfreq,Hinsen:memory} and a self-consistency method.\cite{Tankeshwar:memory} The memory function couched in terms of radial distribution functions has also been studied numerically.\cite{Gaskell:memory} Memory functions thus analyzed have two general characteristics in common, a rapid decay within the first few hundred femtoseconds of relaxation and a long tail thereafter.
The former among these features originates mainly from short-time collisions via fast inertial motions of molecules in liquids.  The residual memory effect which can last for a prolonged period governs the long-time behavior of relaxation dynamics. In solvation dynamics, the overall friction given by the integration of the memory function including its long-time tail is directly related to the solvation time.

In this study, we investigate the power spectra of the FC energy gap in the room-temperature ionic liquid,
1-ethyl-3-methylimidazolium hexafluorophosphate (${\rm EMI}^+{\rm PF}_6^-$), employing a diatomic probe solute via molecular dynamics (MD) computer simulations.   Among others, $1/f$ dependence and crossover to the white noise are observed in the low frequency region of the spectra.  The latter represents the onset of normal diffusion, which marks the escape from the subdiffusive regime in the intermediate time scale corresponding to the $1/f$ region.\cite{Kaneko:diffusion} In the GLE description of solvation dynamics, the memory function can be expressed conveniently
in terms of the power spectrum of the FC energy gap. Using a simple model description for the power spectrum, we compute the memory function and confirm that the crossover frequency determines the time scale of the memory function
decay and is inversely proportional to the solvation time.

This paper is organized as follows: In Sec.~\ref{sec:GLE}, we briefly review solvation dynamics, its GLE description and 
power spectrum analysis. In Sec.~\ref{sec:simulation}, simulation methods are described. MD results for the time correlation functions and power spectra are presented in Sec.~\ref{sec:PSresult}, while the memory functions are analyzed by using a model description in Sec.~\ref{sec:memory}.  
Concluding remarks are offered in Sec.~\ref{sec:conclusion}.

\section{Solvation dynamics}
\label{sec:GLE}

In this section we review briefly the time correlation function of the FC energy gap and its power spectrum, and the GLE approach to equilibrium solvation dynamics. We assume that the probe solute is characterized by an active electronic state $a$ and a reference electronic state $r$ as in previous MD  studies.\cite{shim2,jeong:pol,shim:rtil:review}
For a given solvent configuration, the FC energy  associated with the $a \rightarrow r$ transition of the solute is given by
\begin{equation}
\Delta E_{a\rightarrow r} = E_r - E_a\ ,
\end{equation}
where $E_{a,r}$ denotes the total energy of the solute-solvent system with the solute in state $a$ and $r$, respectively.

We consider the equilibrium solvation dynamics, characterized by the normalized time correlation function
\begin{equation}
C_{a/r}(t)\equiv \frac{\langle \delta\Delta E_{a\rightarrow r}(0)\delta\Delta E_{a\rightarrow r}(t)\rangle}
{\langle (\delta\Delta E_{a \rightarrow r})^2 \rangle}\ ,
\end{equation}
where $\langle \cdots \rangle$ denotes the equilibrium ensemble
average in the presence of the $a$-state solute and $\delta\Delta
E_{a\rightarrow r} \equiv \Delta E_{a\rightarrow r}-\langle \Delta
E_{a \rightarrow r} \rangle$ depending on time $t$. Henceforth the
subscripts representing the solute electronic states are suppressed
for brevity.

The time correlation function is conveniently described by the GLE,\cite{solventfreq,hynes:GLE}
derived via the Mori-Zwanzig projection\cite{MoriZwanzig,Friedman} onto a set of
dynamical variables $\{ \delta \Delta E, \delta \Delta \dot{E}\}$:
\begin{equation}
\delta \Delta \ddot{E}(t)=-\omega_s^2\, \delta \Delta E(t)- \int_0^t dt'\,
\zeta(t-t')\, \delta \Delta \dot{E}(t') + R(t), 
\label{eq:GLE}
\end{equation}
where the solvent frequency $\omega_s \equiv \sqrt{\langle(\delta\Delta \dot{E})^2\rangle\langle(\delta\Delta E)^2\rangle^{-1}}$ characterizes inertial dynamics
of $\Delta E$,\cite{solventfreq} and the time-dependent friction (i.e.,
the memory function) $\zeta(t)$ and the random force $R(t)$ (scaled by the inertia
associated with $\delta \Delta E$ dynamics) 
 are related via the fluctuation-dissipation theorem
\begin{equation}
\zeta(t)=\frac{\langle R(0)R(t)\rangle}{\langle(\delta \Delta \dot{E})^2\rangle}\ .
\end{equation}
The equilibrium time-correlation function $C(t)$ satisfies
\begin{equation}
\ddot{C}(t)=-\omega_s^2 C(t)- \int_0^t dt' \zeta(t-t') \dot{C}(t')\ .
\label{eq:GLEforTCF}
\end{equation}
Taking the Laplace transform
\begin{equation}
\tilde C(z) = \int_0^\infty \exp(-zt)\, C(t) dt\ ,
\end{equation}
we obtain
\begin{equation}
\label{eq:LTofC}
\tilde C(z)=\left( z + \frac{\omega_s^2}{z+\tilde \zeta(z)} \right) ^{-1}\ ,
\end{equation}
which shows that the solvation time
$\tau_{\rm solv}\equiv \int_0^\infty C(t)dt = \tilde C(0)$
is governed by $\tilde \zeta(0)$ and $\omega_s^2$.

The  normalized power spectrum of the FC energy gap $\delta \Delta E_{a \rightarrow r}$ is defined to be
\begin{equation}
S_{a/r}(f) \equiv \lim_{T\rightarrow \infty} \frac{1}{T} \Bigg| \int_{-T/2}^{T/2} \frac{\delta\Delta E_{a \rightarrow r}(t)}{\sqrt{\langle (\delta\Delta E_{a \rightarrow r})^2 \rangle}} e^{2\pi ift}dt \Bigg|^{2}\ ,
\end{equation}
which corresponds to the Fourier transform of the (normalized) correlation function $C(t)$,
according to the well-known Wiener-Khintchine theorem.\cite{Kubo:nonequil}
Namely, the correlation function, which is real, relates to the power spectrum via 
\begin{equation}
C(t)= 2\pi \int_{-\infty}^\infty df S(f) \cos 2\pi ft \ .
\label{eq:WKthm}
\end{equation}
Because of the discrete nature of sampling in simulations, we use the discretized form of
the power spectrum:\cite{NumericalRecipe}
\begin{equation}
S(f_k)=\frac{\Delta t}{N} \Bigg| \sum_{n=0}^{N-1} \frac{\delta\Delta E_n }{\sqrt{\langle (\delta\Delta E)^2 \rangle}} e^{2\pi ink/N} \Bigg|^2, \label{S}
\end{equation}
where $N$ is the number of samples, $f_k \,(\equiv k/N\Delta t)$ is
the $k$th frequency ($k=0,\cdots, N{-}1$) and $\Delta t$ is the sampling interval.  
Eq.~(\ref{eq:WKthm}) shows that the power spectrum $S(f)$ carries all the information relevant to the time correlation function $C(t)$ and elucidates its behavior in the frequency domain. Therefore $S(f)$ can reveal insight into
dynamic character  and time scales of the system.  For example, in the case of Debye relaxation given by single-exponential decay, $S(f)$ is a Lorentzian that decreases as $1/f^2$ at high frequencies.  If the system is characterized by multiple relaxation times, say, $f_1^{-1}$ and $f_2^{-1}$ (with $f_1\ll f_2$), its power spectrum displays a
plateau in the low frequency region ($f\ll f_1 $) and $1/f^2$ decay in the high frequency limit $f \gg f_2$.  In the intermediate region $f_1\ll f \ll f_2$, $S(f)$ varies approximately as $1/f$.\cite{review:1/f}

\section{Simulation Methods}
\label{sec:simulation}

The simulation cell is comprised of a rigid diatomic solute immersed in 
${\rm EMI}^+{\rm PF}_6^-$, consisting of 112 pairs of rigid cations and anions.
We consider two different solute charge distributions:  a neutral pair (NP) without charges and an ion pair (IP) with unit charge separation.  When solvent dynamics occurs in the presence of NP, i.e. NP is the active electronic state, 
IP is regarded as the reference electronic state, and vice versa. The Lennard-Jones (LJ) parameters employed for each constituent atom of the solute are $\sigma= 4$\,{\AA} and $\epsilon/k_B=100$\,K (with the Boltzmann
constant $k_B$).  Its bond length remains fixed at $3.5$\,{\AA}  for all cases considered here.
As for the solvent interaction potential, we employ the same parametrization as in Ref.~\onlinecite{shim1}.

MD simulations were conducted in the canonical ensemble at temperature $T=400$\,K through the use of the DL$\_$POLY program.\cite{dlpoly}  For each solute charge distribution, we simulated the combined solute-solvent system for 70\,ns after 6\,ns equilibration.  To compute the power spectrum of the energy gap with reduced noise, we divided the trajectory into seven 10\,ns segments and calculated $S(f)$ in Eq.~(\ref{S}) by averaging over the seven segments.
The sampling time interval $\Delta t$ in our simulations was 10\,fs, which sets the maximum frequency of our power spectrum analysis at 50\,ps$^{-1}$, while the minimum is $10^{-4}$\,ps$^{-1}$.  

For comparison, we also performed simulations in aprotic acetonitrile  at
$T=300$\,K.  In the simulation cell, a single NP (or IP) solute is immersed in 512 rigid molecules of
acetonitrile. The LJ parameters and partial charges for acetonitrile were taken from
Ref.~\onlinecite{acetonitrile}. The trajectory was 20\,ns long and
the FC energy gap was saved at every 2\,fs.

\section{Time correlation function and power spectrum}
\label{sec:PSresult}
\begin{figure*}
\begin{minipage}{15cm}
\includegraphics[width=7.0cm]{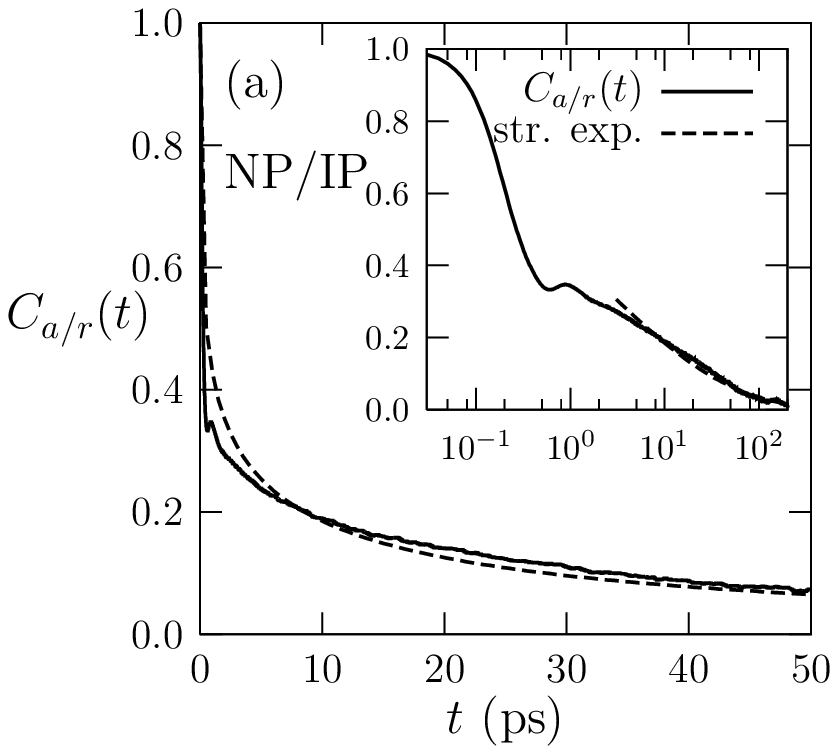}\hfill
\includegraphics[width=7.0cm]{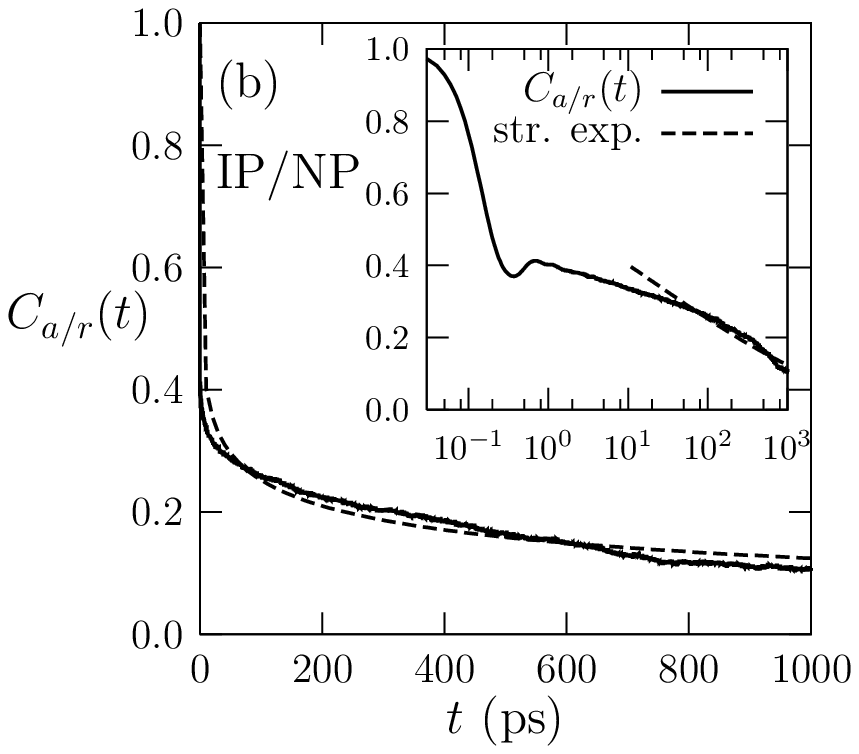}
\end{minipage}\\
\vspace* {25pt}
\begin{minipage}{15cm}
\includegraphics[width=7.0cm]{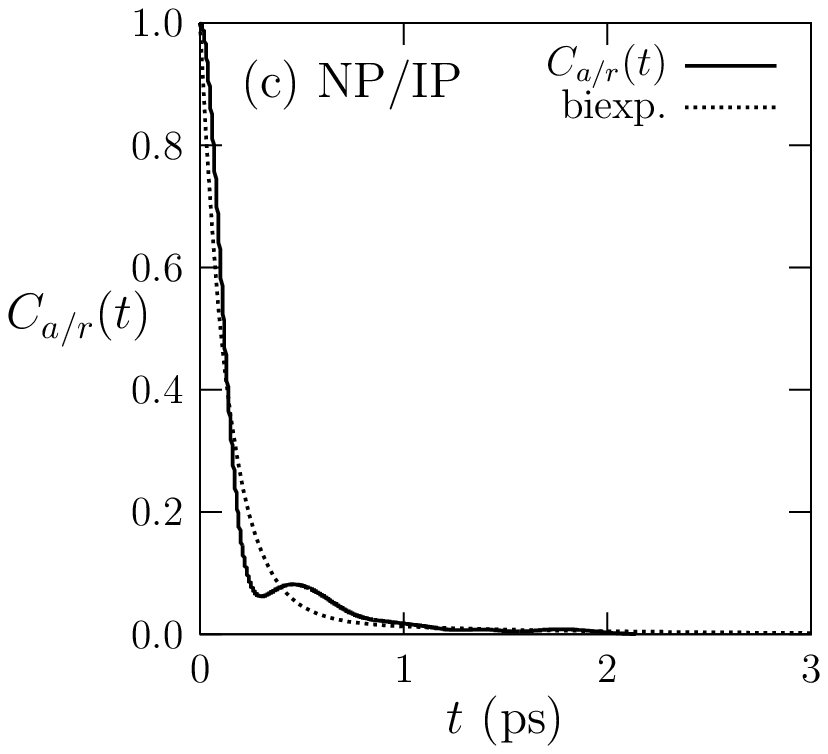}\hfill
\includegraphics[width=7.0cm]{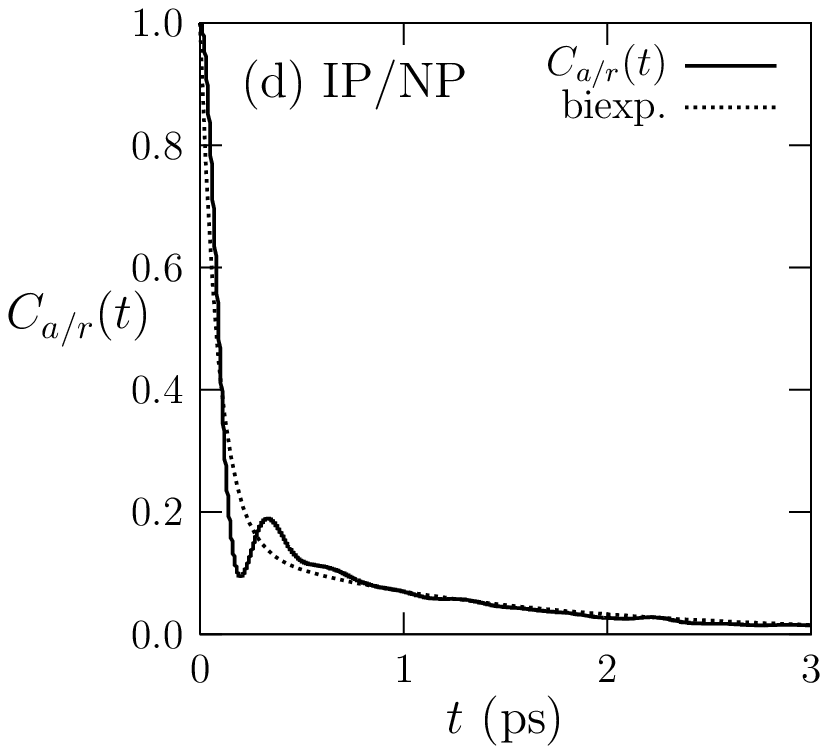}
\end{minipage}\\
\vspace* {35pt}
\caption{Time correlation function $C_{a/r}(t)$ of $\delta\Delta E_{a \rightarrow r}(t)$ in (a) ${\rm EMI}^+{\rm PF}_6^-$ for NP/IP, (b) ${\rm EMI}^+{\rm PF}_6^-$ for IP/NP, (c) ${\rm CH_3CN}$ for NP/IP, and (d) ${\rm CH_3CN}$ for IP/NP active/reference states of the solute. Insets in (a) and (b) display semi-log plots for $C_{a/r}(t)$ }
\label{fig:Ct}
\end{figure*}

In this section, we present MD results for equilibrium solvation dynamics in  ${\rm EMI}^+{\rm PF}_6^-$ and acetonitrile.  
We begin with the time correlation functions of the FC energy gap fluctuations in the presence of the NP and IP solutes  in Fig.~\ref{fig:Ct}.

The MD results for $C_{a/r}(t)$ in ${\rm EMI}^+{\rm PF}_6^-$ exhibit biphasic relaxation, i.e., ultrafast inertial relaxation followed by an extremely slow decay [Fig.~\ref{fig:Ct}(a) and (b)], consonant  with prior simulation studies.\cite{shim:letter,margulis:rtil:dyn,kobrak:rtil:dyn1,shim2,kobrak:rtil:dyn2,jeong:pol,shim:rtil:review}  Long-time behaviors of $C_{a/r}(t)$ are well described by the stretched exponential function $\exp[-(t/\tau_0)^\beta]$:\cite{shim:letter,shim2,jeong:pol}  A good agreement with the MD results was obtained with fitting parameters $\beta=0.30$ and $\tau_0 =1.75\,{\rm ps}$ 
for $1\,{\rm ps} \lesssim t \lesssim 50\,{\rm ps}$ in the case of  NP.  For IP, the stretched exponential fit with $\beta=0.18$ and $\tau_0 =16.84\,{\rm ps}$ applies to a time range, $1\,{\rm ps} \lesssim t \lesssim 1000\,{\rm ps}$, which is about 20 times wider than the NP case. The result that the $\tau_0$ value with IP is larger than that with NP by one order of magnitude is a direct consequence of slow solvent relaxation dynamics in the presence of the former solute, compared with the latter. For instance, the $C_{a/r}(t)$ value reduces to below 0.1 after $t\gtrsim 1\,{\rm ns}$ in the presence of IP, whereas it takes much shorter $\sim 34\,{\rm ps}$ with NP.
The corresponding solvation times are $\tau_{\rm solv} =450$ and 11\,ps for the IP and NP solutes, respectively.

For comparison, we consider $C_{a/r}(t)$ in acetonitrile in Fig.~\ref{fig:Ct}(c) and (d). We notice that solvation dynamics in acetonitrile are much faster than those in ${\rm EMI}^+{\rm PF}_6^-$, congruent with previous studies. The MD results for $\tau_{\rm solv}$ for NP and IP are 0.17 and 0.29\,ps, which are smaller than the ${\rm EMI}^+{\rm PF}_6^-$ values by 2--3 orders of magnitude. The biexponential functions $c\exp(-t/t_1)+(1-c)\exp(-t/t_2)$ 
with $c=0.97$, $t_1=0.14\,{\rm ps}$, $t_2=1.086\,{\rm ps}$ for NP and $c=0.85$, $t_1=0.09\,{\rm ps}$, $t_2=1.31\,{\rm ps}$ for IP provide excellent fits for $C_{a/r}(t)$ in acetonitrile for $t\gtrsim 1$\,ps.  As analyzed below, $S_{a/r}(f)$ associated with $C_{a/r}(t)$ in acetonitrile does not show pronounced $1/f$ behavior.  Thus while a stretched exponential function also appears to yield a reasonable fit (result not shown here), we will take the view that  $C_{a/r}(t)$ in acetonitrile is biexponential with two relaxation times.  In the long-time limit of this description,
$C_{a/r}(t)$ becomes a single exponential decay characterized by the longer of the two relaxation times. 

\begin{figure*}
\begin{minipage}{15.0cm}
\includegraphics[width=7.0cm]{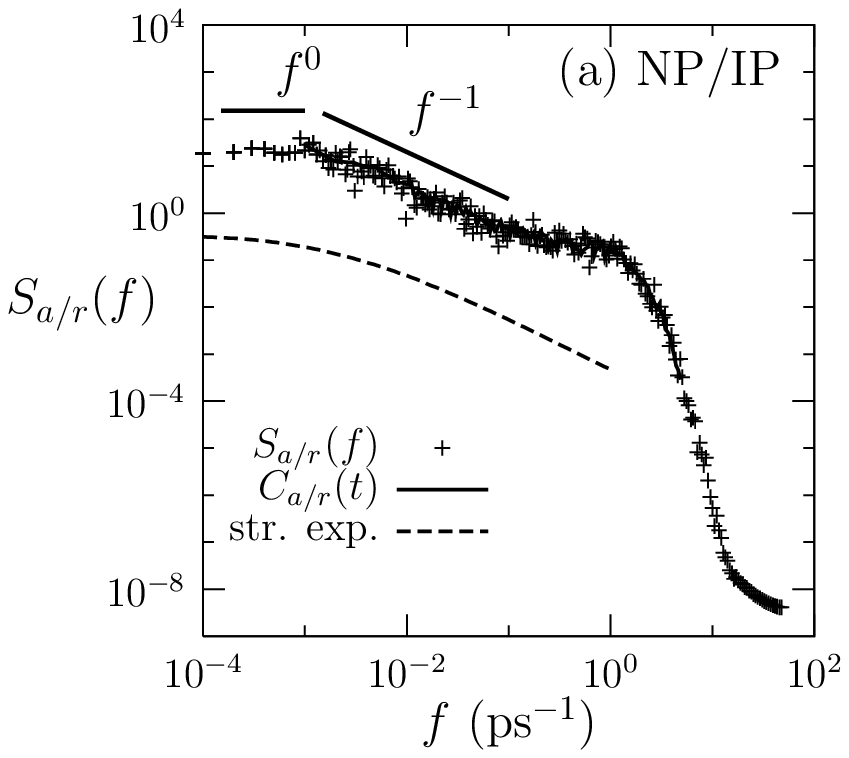}\hfill 
\includegraphics[width=7.0cm]{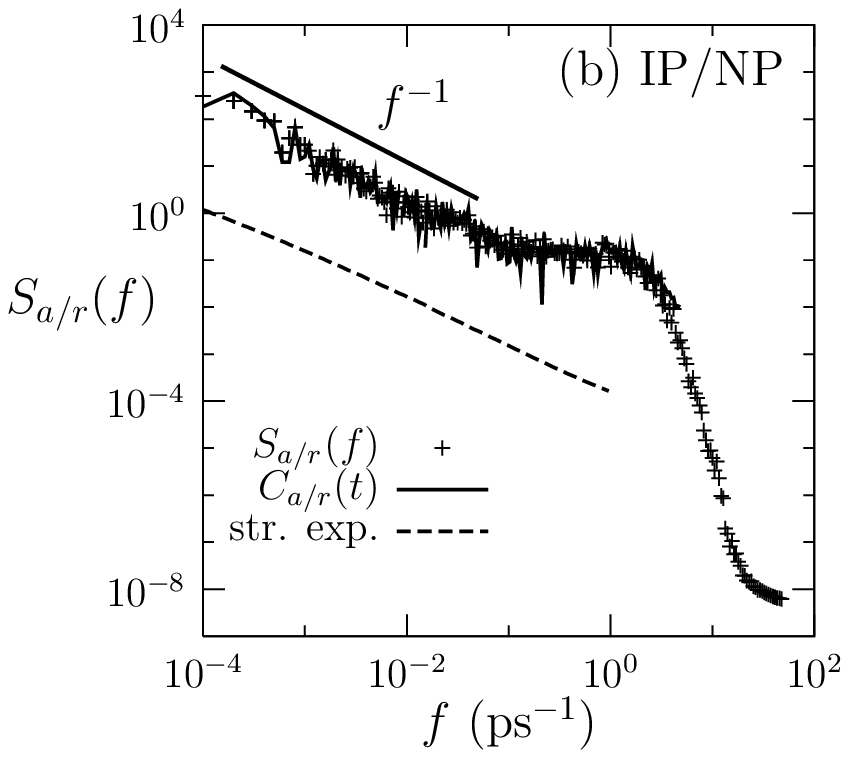}
\end{minipage}\\
\vspace* {25pt}
\begin{minipage}{15.0cm}
\includegraphics[width=7.0cm]{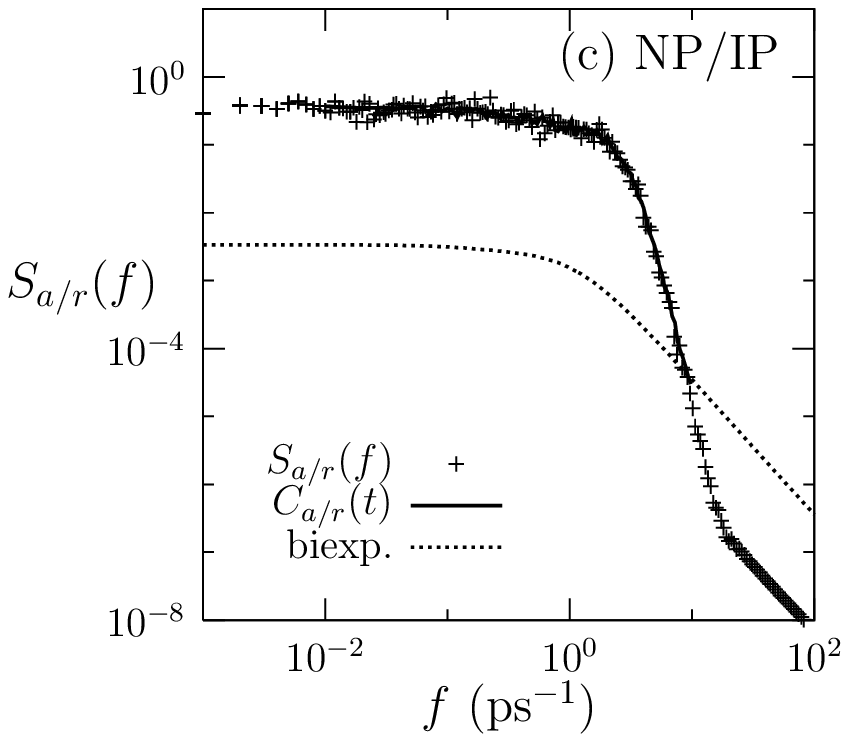}\hfill 
\includegraphics[width=7.0cm]{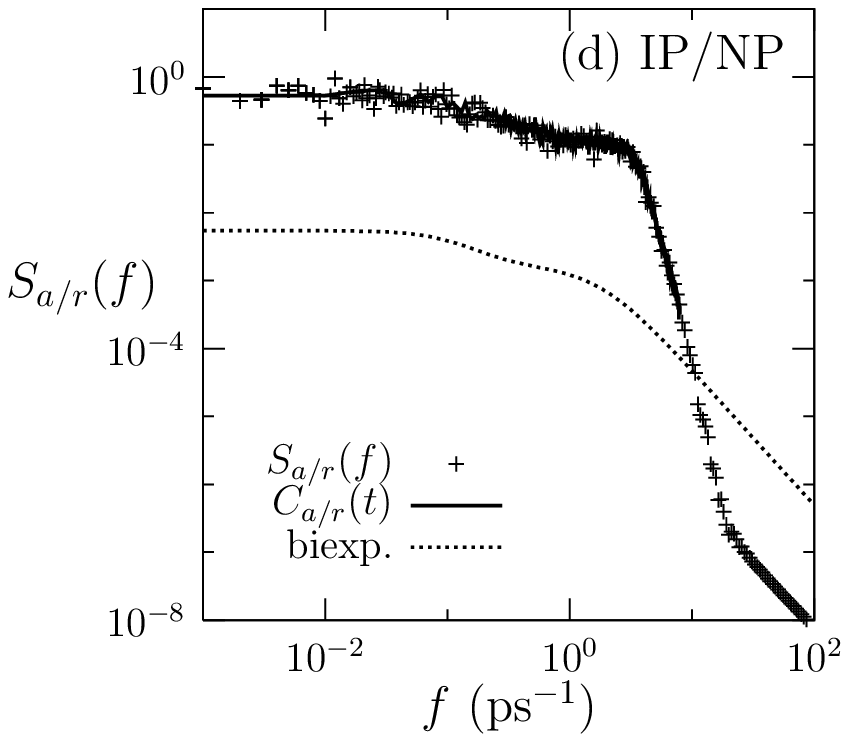}
\end{minipage}\\
\vspace* {35pt}
\caption{Power spectrum 
$S_{a/r}(f)$ of $\delta\Delta E_{a \rightarrow r}(t)$. The results obtained via Eq.~(\ref{S}) are plotted in ``+'' symbols, while the Fourier transforms of $C_{a/r}(t)$ are given in a solid line.  They show an excellent agreement as they should.
Also displayed are Fourier
transforms of the stretched exponential and biexponential fits.  (a), (b), (c), and (d) refer to the same cases
as in Fig.~\ref{fig:Ct}.} 
\label{fig:PS}
\end{figure*}

To gain additional insight into characteristics of equilibrium solvation dynamics, we consider the power spectrum $S_{a/r}(f)$ of the fluctuating FC energy gap $\Delta E_{a \rightarrow r}(t)$.  Figure~\ref{fig:PS} presents $S_{a/r}(f)$  in ${\rm EMI}^+{\rm PF}_6^-$ and in acetonitrile, obtained from MD simulations  with the aid of Eq.~(\ref{S}).  We observe in Fig.~\ref{fig:PS}(a) that $S_{a/r}(f)$ of NP in ${\rm EMI}^+{\rm PF}_6^-$ is characterized by at least four different regimes:\cite{review:1/f,Ramaswamy:water:1/f,Chakravarty:1/f} In the low frequency region below $0.001\,{\rm ps}^{-1}$,  the spectrum is of a white-noise type, which reveals the absence of correlation between two events separated by longer than $\sim1$\,ns.  For later use,  we denote as $f_c$ (``crossover frequency'') the frequency below which $S_{a/r}(f)$ flattens out.  For the NP solute under consideration here, $f_c \approx 10^{-3}$\,ps$^{-1}$. As $f$ increases above $f_c$, the power spectrum exhibits the $1/f$ behavior, which is often interpreted as the presence of many different relaxation time scales.\cite{review:1/f} If the frequency further increases beyond $\sim 1$\,ps$^{-1}$, $S_{a/r}(f)$ begins to drop sharply.  This is attributed to the rapid initial decay of $C_{a/r}(t)$ at short times [cf.\ Fig.~\ref{fig:Ct}(a)].  Finally, in the high frequency region $f\gtrsim 10$\,ps$^{-1}$, the power spectrum decreases as $\sim1/f^2$.  For perspective, the reader is reminded that $f=10^{-4}$\,ps$^{-1}$ and 50\,ps$^{-1}$ are the lower and upper limits of our analysis of the MD results in the frequency domain.

The power spectrum for IP in ${\rm EMI}^+{\rm PF}_6^-$ in Fig.~\ref{fig:PS}(b), though similar to that for NP, shows a couple of interesting differences.  First, we were not able to observe a white-noise spectrum in the low frequency region of $S_{a/r}(f)$ in the presence of IP.   The most likely reason is that the MD trajectory is not long enough to probe 
the onset of the complete loss of correlations because $C_{a/r}(t)$ in the IP case decays much more slowly than that in the NP case.   We ascribe this difference in $C_{a/r}(t)$ to electrostriction.  To be specific, it tends to enhance the solvation structure around IP compared to NP and in turn makes rugged the landscape of the potential energy surface, upon which solvent ions diffuse.\cite{shim:letter} This leads to slower diffusion of solvent ions and therefore slower relaxation of $\Delta E$ fluctuations in the presence of IP than NP. We thus expect that the white-noise plateau for the former solute will appear at frequencies lower than $10^{-4}\, {\rm ps}^{-1}$.  This means that $f_c$ for IP would be lower than $f_c$ for NP, $10^{-3}$\,ps$^{-1}$, by more than one decade.  Second, $S_{a/r}(f)$ with IP shows a flat region for $0.1\,{\rm ps}^{-1} \lesssim f \lesssim 1\,{\rm ps}^{-1}$.  While $S_{a/r}(f)$ with NP also has a hint of a plateau around $f=1$\,ps$^{-1}$, it is much more prominent in the case of IP. The plateau is a part of the Gaussian power spectrum associated with the ultrafast initial relaxation of $C_{a/r}(t)$.  One quick and easy way to see this is that infinitely fast  delta function relaxation yields a plateau in the frequency domain. This plateau is more noticeable in IP because the difference in time scales between its inertial and diffusive dynamics is larger than that of NP.

The power spectra of the stretched exponential fits  in ${\rm EMI}^+{\rm PF}_6^-$ are compared with the MD results in Fig.~\ref{fig:PS}(a) and (b).  For both IP and NP, the stretched exponential functions well describe the $1/f$ characteristics---both the range and exponent of the power law behavior---of the simulation results. One prevalent notion is that stretched exponential behavior arises from the superposition of different single exponential decays, weighted by a broad distribution of relaxation times.\cite{Richter:KWW}  While the assumption of single exponentials per se may be too restrictive, there is considerable evidence that RTIL dynamics are characterized by a distribution of different time scales.\cite{voth:rtil:hetero,ribeiro:rtil:hetero,margulis:rtil:ree,shim:rtil:rot,maginn:rtil3} As pointed out in Sec.~\ref{sec:GLE}, the power spectrum of the system, involving a number of relaxation processes of different time scales, in general exhibits $1/f$ dependence.\cite{review:1/f} Thus the presence of the $1/f$-type domain in the power spectrum of $\Delta E$ is another manifestation of non-exponential RTIL relaxation. This also suggests the similarity between solvation dynamics in the RTILs and the dynamic heterogeneity observed in glassy liquids.\cite{Richert:review:hetero,Ediger:review:glass,Colmenero:KWW}

Before we turn to memory functions, we briefly consider $S_{a/r}(f)$ for acetonitrile in Figure~\ref{fig:PS}(c) and (d). In the case of NP, $S_{a/r}(f)$  displays a white-noise plateau for $f \lesssim 2\,{\rm ps}^{-1}$ but it does not show a $1/f$ power-law behavior. For IP, the white-noise spectrum obtains for $f \lesssim 0.08\,{\rm ps}^{-1}$.  As $f$ increases beyond the white-noise region, $S_{a/r}(f)$ briefly shows $1/f^\alpha$ behavior with $\alpha \simeq 0.53$ before it begins a rapid decrease. As in ${\rm EMI}^+{\rm PF}_6^-$, the crossover frequency $f_c$ for IP is smaller than that for NP, again due to electrostriction. Regardless of the solute charge distributions, the onset of the white-noise region is at much higher frequencies in acetonitrile  than in ${\rm EMI}^+{\rm PF}_6^-$ because the long-time solvent relaxation in the former is much faster than that in the latter.  Furthermore, the $1/f$ spectrum is nearly absent in $S_{a/r}(f)$ of acetonitrile in contrast to the ${\rm EMI}^+{\rm PF}_6^-$ case.  As mentioned above, this is why we favor a biexponential description for $C_{a/r}(t)$ in acetonitrile over a stretched exponential description. We notice that except for the rapid decrease in the neighborhood of $f=10$\,ps$^{-1}$ arising from ultrafast inertial relaxation of $C_{a/r}(t)$, the biexponential functions indeed provide an excellent framework to describe $S_{a/r}(f)$ in acetonitrile. 

\section{Memory function analysis}
\label{sec:memory}

Here we consider a simple model description for $S(f)$ and analyze its memory function $\zeta(t)$ to gain insight into long-time solvation dynamics. As observed above, power spectra in ${\rm EMI}^+{\rm PF}_6^-$ are characterized by at least four different regimes; viz.,  $S(f)$ generally show white-noise, $1/f$, sharp fall-off and $1/f^2$ behaviors in turn as the frequency increases. Since we are mainly interested in long-time dynamics, we ignore the frequency region (i.e., $1\,{\rm ps}^{-1}\lesssim f \lesssim 10$\,ps$^{-1}$) of rapid $S(f)$ decay, which arises mainly from ultrafast inertial relaxation of $C(t)$. We thus consider the power spectrum of the following form: 
\begin{equation}
S(\omega)= \left\{ \begin{array}{ll}
A\omega_c^{-1} & \textrm{for $0<\omega<\omega_c$\ ,}\\
A \omega^{-1} & \textrm{for $\omega_c<\omega<\omega_0$\ ,}\\
A\omega_0\omega^{-2} & \textrm{for $\omega_0<\omega<\omega_m$\ ,}\\
0 & \textrm{for $\omega>\omega_m$\ ,}
\end{array}\right.
\label{eq:SD}
\end{equation}
where $S(-\omega)=S(\omega)$, $\omega \equiv 2\pi f$ and $A$ is the normalization constant given by
$2A\equiv \left[2-\omega_0 /\omega_m +\ln (\omega_0 /\omega_c)\right]^{-1}$.
By differentiating Eq.~(\ref{eq:WKthm}) twice with respect to $t$ and setting $t=0$,
we derive a sum rule\cite{Friedman}
\begin{equation}
\label{eq:sumrule}
\omega_s^2=\int_{-\infty}^{\infty} d\omega\,\omega^2 S(\omega)
=\frac{\omega_0\omega_m-\omega_0^2/2-\omega_c^2/6}
                {2-\omega_0/\omega_m+\ln(\omega_0/\omega_c)},
\end{equation}
which clearly shows the necessity of a high frequency cutoff at $\omega_m$ in Eq.~(\ref{eq:SD}).

The parameters employed to model $S(f)$ are compiled in Table~\ref{table:freq}.  S1 and S2 there are the model descriptions for solvation of NP and IP in ${\rm EMI}^+{\rm PF}_6^-$, respectively, while the corresponding cases in acetonitrile are modeled by S3 and S4.   Thus S1 to S4 correspond to the cases in Fig.~\ref{fig:PS}(a) to (d), respectively.
The solvent frequency $\omega_s$ was obtained from the simulations via Eq.~(\ref{eq:GLE}) and the cutoff frequency $\omega_m$ was estimated from Eq.~(\ref{eq:sumrule}).  The values of $\omega_c$ and $\omega_0$ were determined with the aid of the MD results in Fig.~\ref{fig:PS}.  In the presence of IP in ${\rm EMI}^+{\rm PF}_6^-$, we  were not able to obtain $\omega_c$ from the simulation because
the plateau behavior in the low frequency region of $S(f)$ was not accessible as mentioned above [Fig.~\ref{fig:PS}(b)].
In the absence of any additional information on the crossover frequency, we assumed $\omega_c = 2\pi\times 10^{-4}\,{\rm ps}^{-1}$, the lowest frequency that is allowed in our analysis of MD results, for S2. The corresponding crossover frequency for S1, viz., NP in ${\rm EMI}^+{\rm PF}_6^-$, is $\omega_c \approx 2\pi\times 10^{-3}\,{\rm ps}^{-1}$.   For S3, we employed $\omega_c =\omega_0$ because acetonitrile does not exhibit a $1/f$ spectrum in the presence of NP.
The narrow $1/f$ region present in S4 was incorporated into the model calculations by choosing
$\omega_c=0.16\pi$\,ps$^{-1}$ and $\omega_0=4\pi$\,ps$^{-1}$.

Within the model description of Eq.~(\ref{eq:SD}), we can determine $\zeta(t)$ exactly.   Specifically, we make Laplace transform of Eq.~(\ref{eq:WKthm}) with  Eq.~(\ref{eq:SD}) to find
\begin{eqnarray}
\label{eq:Cz}
\tilde{C}(z) &=& \int_{-\infty}^{\infty}d\omega S(\omega)\frac{z}{\omega^2+z^2} \nonumber \\
&=& 2A\left[\int_0^{\omega_c}d\omega \frac{z}{\omega_c(\omega^2+z^2)}+
         \int_{\omega_c}^{\omega_0}d\omega\frac{z}{\omega(\omega^2+z^2)}\right. \nonumber \\
& &~~~~\left.  +\int_{\omega_0}^{\omega_m} d\omega \frac{\omega_0 z}
                   {\omega^2 (\omega^2+z^2)}\right] \nonumber\\
&=& 2A\left[\frac{1}{\omega_c}\tan^{-1}\frac{\omega_c}{z} + \frac{1}{2z}\left(\ln\frac{\omega_0^2}{\omega_0^2+z^2}  \right.\right. \nonumber \\
& &~~~~\left. -\ln\frac{\omega_c^2}{\omega_c^2+z^2}\right)
+\frac{\omega_0}{z}\left(\frac{1}{\omega_0}-\frac{1}{\omega_m}\right) \nonumber\\
 & &~~~~ \left. +\frac{\omega_0}{z^2}\left(\tan^{-1}\frac{\omega_0}{z}-\tan^{-1}\frac{\omega_m}{z}\right) \right] \ .
\end{eqnarray}
We rewrite Eq.~(\ref{eq:LTofC}) as
\begin{equation}
\tilde\zeta(z)={\omega_s^2}\frac{\tilde C(z)}{1-z\tilde C(z)} -z \ ,
\label{eq:z+zeta}
\end{equation}
substitute Eq.~(\ref{eq:Cz}) into Eq.~(\ref{eq:z+zeta}) and make inverse Laplace transform\cite{lapin} numerically to
obtain $\zeta(t)$. 

\begin{figure*}
\begin{minipage}{15.0cm}
\includegraphics[width=7.0cm]{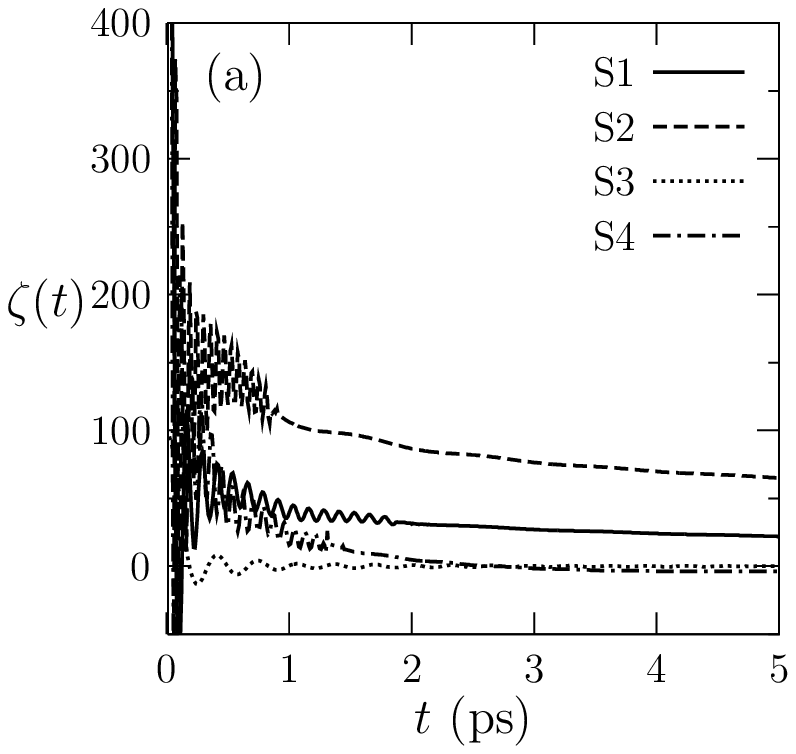} \hfill
\includegraphics[width=7.0cm]{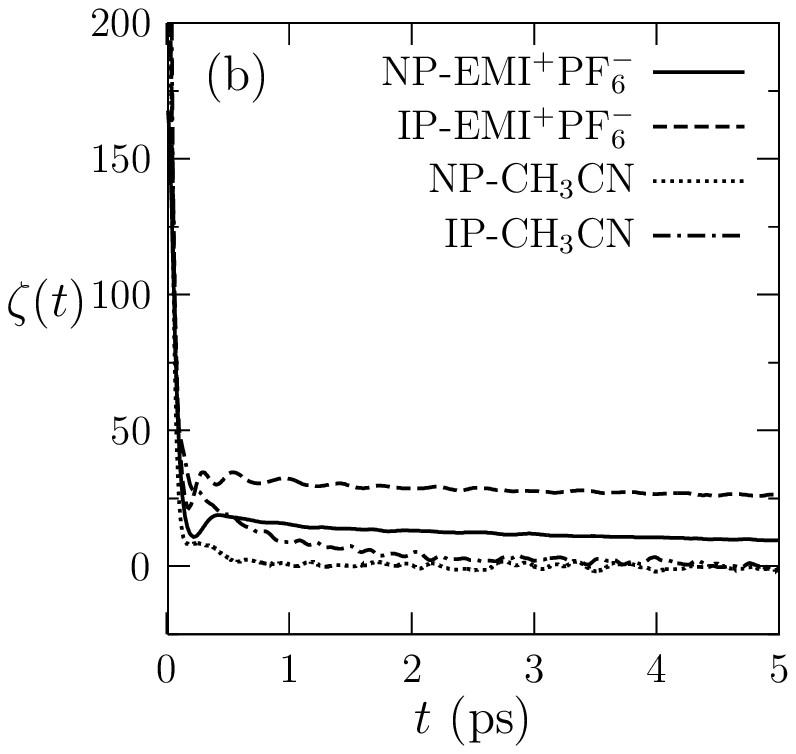}
\end{minipage}\\
\vspace* {35pt}
\caption{
Memory function $\zeta(t)$, (a) obtained via the inverse transform of Eq. (\ref{eq:z+zeta}), 
together with Eq. (\ref{eq:Cz}) and the parameters for S1 to S4 in Table~\ref{table:freq}; 
(b) evaluated directly from the correlation function for the NP and IP solutes in 
${\rm EMI}^+{\rm PF}_6^-$ and in acetonitrile.
}
\label{fig:mem}
\end{figure*}

Figure~\ref{fig:mem}(a) displays the results for the memory function $\zeta(t)$ thus 
obtained.  For comparison, we employ the method used in Ref.~\onlinecite{shim2} to determine $\zeta(t)$ directly from $C(t)$ and associated time correlation functions of a nonconservative force\cite{note:zeta} and present the results in Figure~\ref{fig:mem}(b).
We notice in Figure~\ref{fig:mem}(a) that $\zeta(t)$ falls off very quickly at the very early stage of relaxation,
regardless of the crossover frequency $\omega_c$.
Subsequent decay of the residual memory varies strongly with $\omega_c$.  Generally,
the long-time relaxation of $\zeta(t)$ becomes slower with decreasing $\omega_c$.\cite{note}
Comparison of the results in Figure~\ref{fig:mem}(a) and (b) shows that while there are differences,
$\zeta(t)$ obtained from model $S(\omega)$ in Eq.~(\ref{eq:SD}) correctly captures important  features of memory both at the qualitative and semi-quantitative level.  These include rapid initial decay of a large amplitude and the existence of a long-time tail.  Nonetheless, the model calculations yield spurious oscillations at short times (see below) and overestimation of the long-time memory effect.  For example, the model predictions for the magnitude of the residual memory, say at $t\approx 1$\,ps, are considerably larger than the MD results.   This overestimation is attributed mainly to the neglect of the rapid decay of $S(f)$ in our model description at high frequencies.

\begin{figure*}
\includegraphics[width=7.0cm]{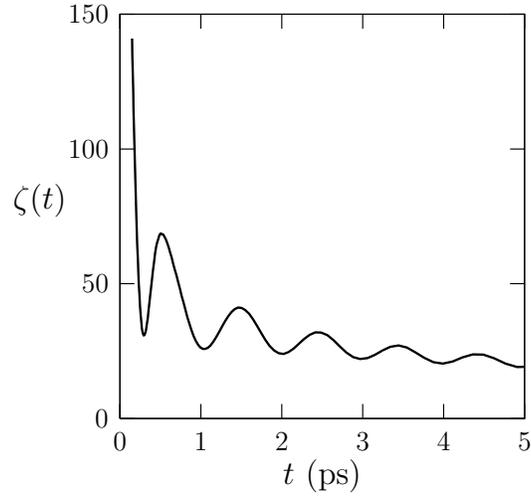}
\vspace* {35pt}
\caption{Memory function $\zeta(t)$ in case that the power spectrum at high frequencies
($\omega>\omega_0$) follows a Gaussian function: 
$S(\omega)= (A/\omega_0) \exp[-(\omega-\omega_0)^2/2\omega_g^2]$
with $\omega_0=2\pi\,{\rm ps^{-1}}$ and $\omega_g$ set to be $0.8\,\omega_0$.
The solvent frequency is given by
$\omega_s=5.7\,{\rm ps^{-1}}$ for $\omega_c=2\pi\times 10^{-3}\,{\rm ps^{-1}}$
according to the sum rule.}
\label{fig:erfc}
\end{figure*}

Here we briefly examine the oscillatory behavior of $\zeta(t)$ observed in the first few hundred femtoseconds
in Fig.~\ref{fig:mem}(a).  It is of interest to note that the period of these oscillations is close to $2\pi/\omega_m$.
This suggests that oscillations are closely linked to the presence of a cutoff at frequency $\omega_m$
in the model power spectrum employed in this study.  To check this, we considered a Gaussian power spectrum $S(f)=(A/\omega_0) \exp[-(\omega-\omega_0)^2/2\omega_g^2]$ for $\omega>\omega_0$ that does not require a high-frequency cutoff.  The resulting friction is exhibited in Fig.~\ref{fig:erfc}. The disappearance of the rapid initial oscillations confirms that indeed the cutoff is mainly responsible for rapid oscillations in $\zeta(t)$ in Fig.~\ref{fig:mem}(a).  We note that slow oscillations in Fig.~\ref{fig:erfc} arise from the discontinuity in the derivative of the model Gaussian spectrum at $\omega=\omega_0$.

Finally, we consider solvation time $\tau_{\rm solv} (=\tilde C(0))$.  It is related to the total memory $\tilde \zeta(0)$ via Eq.~(\ref{eq:LTofC}), so that
\begin{equation}
\frac{\tilde\zeta(0)}{\omega_s^2}=\tau_{\rm solv}=\frac{\pi}{2\omega_c}\left(2-\frac{\omega_0}{\omega_m}+
\ln\frac{\omega_0}{\omega_c}\right)^{-1}\ ,
\label{eq:tau}
\end{equation}
where we have used Eq.~(\ref{eq:Cz}) in passage to the final expression. The results for $\tau_{\rm solv}$ obtained from Eq.~(\ref{eq:tau}) are presented  in Table~\ref{table:freq}.  We notice that $\tau_{\rm solv}$ is inversely proportional to the crossover frequency $\omega_c$.  Thus, all other things being equal, $\tau_{\rm solv}$ increases as $\omega_c$ decreases.  This is directly related to the observation made above that long-time relaxation of $\zeta(t)$ becomes slower with decreasing $\omega_c$. Also interesting is that  $\tau_{\rm solv}$ varies with the frequency range ${\omega_0}/{\omega_c}$ associated with the $1/f$ behavior in $S(f)$.  For instance, for given $\omega_c$, the growing range of the $1/f$ region, i.e., increasing $\omega_0$, tends to reduce the solvation time.  Since the lower limit of distributed time scales for $1/f$ is $\sim\! \omega_0^{-1}$ [cf.\ Eq.~(\ref{eq:SD})], the shortest time scale for $1/f$ becomes faster as $\omega_0$ increases.  In other words, faster processes become available increasingly more to the system, while the availability of slower processes remains unchanged.  This yields the reduction in the solvation time because contributions to solvent relaxation from faster processes become progressively more important than slower processes.

\section{Conclusions}
\label{sec:conclusion}

We have studied the time correlation functions and power spectra of the FC energy gap
and related memory functions associated with solvation dynamics of model diatomic solutes in
${\rm EMI}^+{\rm PF}_6^-$ and ${\rm CH_3CN}$. It was found that the power spectra of both the NP and IP solutes in ${\rm EMI}^+{\rm PF}_6^-$ display $1/f$ dependence over a range of intermediate frequencies, which indicates non-exponential relaxation dynamics in RTILs.  In the case of NP, we observed white-noise behavior at low frequencies.  The power spectrum of NP in acetonitrile does not yield the $1/f$ dependence and its crossover frequency is
much higher than the corresponding value in ${\rm EMI}^+{\rm PF}_6^-$.   Though not pronounced, the IP solute in acetonitrile shows $1/f$ dependence but  in a frequency range much narrower than that  in ${\rm EMI}^+{\rm PF}_6^-$.

Using a simple but analytic model for power spectra, we have determined the memory function in the GLE description of solvation dynamics and compared with the MD results.  With proper account of white-noise, $1/f$ and $1/f^2$ behaviors of power spectra, we have found that the model description reproduces the MD results of friction reasonably well.
We have also found that the time scale of memory effects and the $1/f$ regime are closely related. We have obtained the solvation time in terms of the frequency parameters of the model power spectrum description.  Among others, the solvation time was found to be inversely proportional to the crossover frequency.  Together with our MD results of $S_{a/r}(f)$, this indicates that the solvation time for the IP solute in ${\rm EMI}^+{\rm PF}_6^-$ is far longer than that for NP, at least by one order of magnitude.  Electrostriction which exerts a strong influence on the landscape of the potential energy surface relevant to solvent ion diffusion is mainly responsible for the variation of solvation time with the solute charge distribution.

\section*{Acknowledgments}
This work was supported in part by MOE through the BK21 Program,  by MOST/KOSEF through National Core Research Center for Systems Bio-Dynamics and Science Research Center for Space-Time Molecular Dynamics (R11-2007-012-03003-0), and by Research Settlement Fund for the new faculty of SNU.

\bibliography{ref2}

\begin{thebibliography}{47}
\bibitem[{not({\natexlab{a}})}]{note:zeta}
\bibinfo{note}{In the numerical inversion of Eq. (14) to the time domain, the
  linear term in $z$ should be cancelled out by asymptotic behavior of
  $\omega_s^2 \tilde C(z)/[1-\tilde C(z)]$ for large $z$. Unlike the analytic
  expression in Eq. (13), however, numerical results of $\tilde C(z)$ do not
  satisfy this condition. We have thus determined $\tilde\zeta(z)$ by combining
  small and large $z$ behaviors obtained from $C(t)$ and from the time
  correlation function of the nonconservative force, respectively.}
\bibitem[{not({\natexlab{b}})}]{note}
\bibinfo{note}{We have performed additional calculations by varying $\omega_c$
  with $\omega_0$ and $\omega_m$ fixed and confirmed that $\zeta(t)$ decays
  more slowly with decreasing $\omega_c$.}
\end{thebibliography}

\newpage
\begin{table*}[!t]
\begin{tabular}{c|cccccc}
\hline
   & $\omega_s $ & $\omega_c/2\pi$ & $\omega_0/2\pi$ & $\omega_m/2\pi$ & $\tau_{\rm solv}$
   & Solute-solvent \\
\hline
S1  &  5.82 & $10^{-3}$ &  1 & 8.04 & 28.4   &NP-${\rm EMI}^+{\rm PF}_6^-$ \\
S2  &  7.81 & $10^{-4}$ &  1 & 17.7 & 224.1  &IP-${\rm EMI}^+{\rm PF}_6^-$ \\
S3  &  10.23 & $2$      &  2 & 3.14 & 0.07   &NP-${\rm CH}_3{\rm CN}$      \\
S4  &  12.99 & $0.08$   &  2 & 11.8 & 0.61   &IP-${\rm CH}_3{\rm CN}$      \\
\hline
\end{tabular}
\vspace*{35pt}
\caption{Frequency parameters for the power spectrum in Eq.~(\ref{eq:SD}) and
the solvation time $\tau_{\rm solv}$ given by Eq. (\ref{eq:tau}). The frequency and
time are measured in the units of ${\rm ps}^{-1}$ and ps, respectively.
S1 through S4 model the power spectra obtained from MD simulations of solute-solvent 
systems shown in the last column.
}
\label{table:freq}
\end{table*}

\end{document}